\documentclass[
 reprint,
 amsmath,amssymb,
 aps,
]{revtex4-2}

\usepackage{graphicx}
\usepackage{dcolumn}
\usepackage{bm}
\usepackage{xcolor}
\usepackage{hyperref}
\usepackage[normalem]{ulem}
\usepackage{xtab,afterpage,longtable}

\hypersetup{
colorlinks=true,
citecolor=blue,
linkcolor=blue,
urlcolor=blue}

\begin{document}

\preprint{APS/123-QED}

\title{Diffuse supernova neutrino background with up-to-date star formation rate measurements and long-term multidimensional supernova simulations}

\author{Nick Ekanger$^1$}
\email{enick1@vt.edu}
\author{Shunsaku Horiuchi$^{1,2}$}
\email{horiuchi@vt.edu}
\author{Hiroki Nagakura$^{3}$}
\author{Samantha Reitz$^{1,4}$}
\affiliation{${}^1$Center for Neutrino Physics, Department of Physics, Virginia Tech, Blacksburg, Virginia 24061, USA.}
\affiliation{${}^2$Kavli IPMU (WPI), UTIAS, The University of Tokyo, Kashiwa, Chiba 277-8583, Japan}
\affiliation{${}^3$Division of Science, National Astronomical Observatory of Japan, 2-21-1 Osawa, Mitaka, Tokyo 181-8588, Japan}
\affiliation{${}^4$Department of Physics, Radford University, Radford, Virginia 24142, USA}

\date{\today}

\begin{abstract}
The sensitivity of current and future neutrino detectors like Super-Kamiokande (SK), JUNO, Hyper-Kamiokande (HK), and DUNE is expected to allow for the detection of the diffuse supernova neutrino background (DSNB). However, the DSNB model ingredients like the core-collapse supernova (CCSN) rate, neutrino emission spectra, and the fraction of failed supernovae are not precisely known. We quantify the uncertainty on each of these ingredients by (\textit{i}) compiling a large database of recent star formation rate density measurements, (\textit{ii})  combining neutrino emission from long-term axisymmetric CCSNe simulations and strategies for estimating the emission from the protoneutron star cooling phase, and (\textit{iii})  assuming different models of failed supernovae. Finally, we calculate the fluxes and event rates at multiple experiments and perform a simplified statistical estimate of the time required to significantly detect the DSNB at SK with the gadolinium upgrade and JUNO. Our fiducial model predicts a flux of $5.1\pm0.4^{+0.0+0.5}_{-2.0-2.7}\,{\rm cm^{-2}~s^{-1}}$ at SK employing Gd-tagging, or $3.6\pm0.3^{+0.0+0.8}_{-1.6-1.9}$ events per year, where the errors represent our uncertainty from star formation rate density measurements, uncertainty in neutrino emission, and uncertainty in the failed supernova scenario. In this fiducial calculation, we could see a $3\sigma$ detection by $\sim2030$ with SK-Gd and a $5\sigma$ detection by $\sim2035$ with a joint SK-Gd/JUNO analysis, but background reduction remains crucial.
\end{abstract}

\maketitle

\section{\label{sec:intro}Introduction}

The diffuse supernova neutrino background (DSNB) is the background of neutrinos from all past stellar core collapse, which occur as the final stage of massive stars with mass above $\sim8\,M_{\odot}$ and each release $\sim10^{58}$ neutrinos \cite{Kotake:2005zn,2016NCimR..39....1M,2017janka,Horiuchi:2018ofe,2021burrows}. These core collapse supernovae, distributed over cosmological timescales, give rise to an isotropic signal of $\sim10\,{\rm MeV}$ neutrinos. The DSNB offers an immediate opportunity to detect core-collapse neutrinos, which in turn offer probes of the historical core-collapse rate, core-collapse neutrino emissions, neutrino physics, and a wide range of beyond Standard Model physics; see, e.g., Refs.~\cite{2010beacom,2016lunardini,2020vitigliano,Suliga:2022ica,Ando:2023fcc} for DSNB reviews. 

Although the DSNB has not been detected yet, the prospect for detection with a wealth of detectors---current and upcoming---is positive. For electron antineutrinos via inverse beta decay (IBD) interactions, there are several existing and upcoming experiments. In Ref.~\cite{Super-Kamiokande:2021jaq}, upper flux limits of the DSNB were placed by Super-Kamiokande (SK) using over eight years of data. Intriguingly this began probing the most optimistic theory predictions. Recently, the next phase of SK began when 0.01\% in mass concentration of gadolinium (SK-Gd) was added (see also Refs.~\cite{Beacom:2003nk,Super-Kamiokande:2021the}) and, only after $\sim552\,{\rm days}$, competitive upper limits were also placed \cite{Super-Kamiokande:2023xup}. Currently, SK-Gd is running with even more Gd, 0.03\% by mass. The Jiangmen Underground Neutrino Observatory (JUNO) is expected to start taking data in 2023 and will collect data alongside SK, searching for $\overline{\nu}_e$ neutrinos from the DSNB \cite{JUNO:2015zny} that can increase the number of detections. The successor to SK, Hyper-Kamiokande (HK), currently under construction and estimated to be completed in 2027, will be $\sim8$ times larger in volume than SK \cite{Hyper-Kamiokande:2018ofw}. In the electron neutrino flavor, the Deep Underground Neutrino Experiment (DUNE) \cite{DUNE:2015lol} is powerful complement to IBD detectors.

\begin{figure*}
\includegraphics[width=\linewidth]{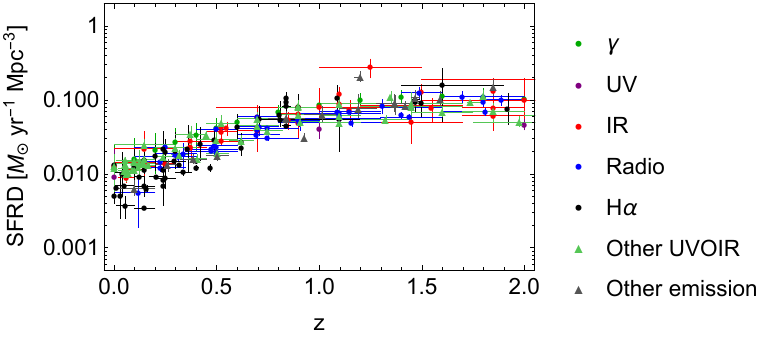}
\caption{Our compilation of recent (post-2006, Refs.~\cite{Takahashi:2007ia,Morioka:2008ra,Shioya:2007kx,Villar:2007bx,Dunne:2008tk,Shim_2009,Smolcic:2008br,sobral2009,Zhu:2008ee,dale2010,rodighiero2010,vaccari2010,westra2010,Bayliss2011,bothwell2011,karim2011,ly2011,robotham2011,cucciati2012,sobral2012,Burgarella:2013jaa,Ciardullo:2013yka,Gunawardhana:2013fha,Gruppioni:2013jna,Magnelli:2013bya,Sobral:2013fga,Planck:2013qqi,Gruppioni:2015bla,khostovan2015,marchetti2016,Sobral:2015tla,davies2016,guijarro2016,delgado2016,rr2016,vansistine2016,Driver:2017qne,novak2017,audcentross2018,coughlin2018,Fermi-LAT:2018lqt,fernandez2018,sanchez2019,upjohn_brown_hopkins_bonne_2019,bellstedt2020,Gruppioni:2020vue,khostovan2020,vilellarojo2021,Yan:2022ctq,cochrane2023}) SFRD measurements up to redshift $z=2$. With different colors and marker styles, we show the indicator used to measure SFRD. All measurements are calibrated to a Chabrier IMF \cite{Chabrier:2003ki} and have been rescaled assuming $H_0=70\,{\rm km~s^{-1}~Mpc^{-1}}$ (see text). The measured values differ only by a factor of a few and, overall, largely agree within uncertainties.}
\label{fig:sfrdindicators}
\end{figure*}

To accurately model the DSNB signal, one needs to understand the rate of core collapse as a function of redshift, the neutrino emission from typical core collapse (i.e., the total energy released and emission spectrum), and the detector response for experiments seeking DSNB detection. Theoretical estimates of the DSNB were initially uncertain by an overall factor of $\sim10$ or more, but this factor has steadily decreased (see Refs.~\cite{Krauss:1983zn,PhysRevLett.55.1422,Totani:1995rg,1996ApJ...460..303T,Malaney:1996ar,Hartmann:1997qe,Kaplinghat:1999xi,Ando:2002ky,Fukugita:2002qw,Strigari:2003ig,Iocco:2004wd,Strigari:2005hu,Lunardini:2005jf,PhysRevD.72.103007,PhysRevC.74.015803,Horiuchi:2008jz,Lunardini:2009ya,PhysRevD.81.083001,Galais:2009wi,PhysRevD.85.043011,Vissani:2011kx,Lunardini:2012ne,2013nakazatoimprint,Mathews:2014qba,Yuksel:2012zy,Nakazato:2015rya,Hidaka:2016zei,2017priya,Horiuchi:2017qja,2018moller,Singh:2020tmt,Kresse:2020nto,Tabrizi:2020vmo,Horiuchi:2020jnc,Ekanger:2022neg,Ashida:2022nnv,Ashida:2023heb} for theoretical models of the DSNB). More specifically, the use of the measured cosmic star formation rate density (SFRD) as an indicator for the core-collapse rate, and the realization of large sets of core-collapse simulations, have been major contributors. However, challenges remain; in particular, core-collapse simulations typically only extend to $\sim 1$ second, while the DSNB requires the neutrino emission time-integrated, ideally to 5--10 seconds \cite{Ekanger:2022neg}. Similarly, there have been dozens of new SFRD measurements in the decade since widely used compilations used in DSNB estimates (e.g., \cite{Hopkins:2006bw,Madau:2014bja}), and a reevaluation of uncertainties is warranted.

The purpose of this work is to incorporate the current state of understanding and uncertainty for these theoretical model ingredients. First and foremost, we compile an extensive list of recent SFRD measurements with various indicators to estimate the rate of core collapse and its error. We combine this with the suite of long-term axisymmetric core-collapse simulations of Ref.~\cite{Nagakura:2021lma}, augmented with analytic strategies to estimate the late phase of average neutrino emission spectrum following Ref.~\cite{Ekanger:2022neg}. 
Finally, we include neutrino emission from failed supernova models \cite{Nakazato:2021gfi,Walk:2019miz} and look to recent studies regarding the fraction of failed supernovae (see the references in Sec.~\ref{subsec:failed}) to understand distinguishing the contribution of successful and failed supernova channels affect the DSNB. These allow us to make our best estimates of the DSNB rate and characterize the uncertainty on these rates.

We organize this study as follows. In Sec.~\ref{subsec:sfrd}, we describe and show recent measurements of the SFRD as a function of redshift in order to infer the cosmic rate of stellar core collapse, which we compare to direct measurements of CCSNe in Sec.~\ref{subsec:directcc}. In Sec.~\ref{sec:neutrinoemission}, we describe the models we use for the neutrino emission from CCSNe, both successful and failed. In Sec.~\ref{subsec:calculations}, we describe the formalism for estimating the rate of DSNB events, quantify how much uncertainty we can expect from ingredients like SFRD measurements, neutrino emission, and the fraction of failed supernovae. We then use this DSNB model in Sec.~\ref{subsec:significantdetection} to estimate their detectability. Finally, we discuss and conclude our results in Sec.~\ref{sec:discussion}.

\section{\label{sec:rcc}Rate of Core Collapse}

\subsection{\label{subsec:sfrd}SFRD measurements}

In order to calculate the DSNB rate, we need to understand how the rate of core collapse, $R_{\rm CC}$, evolves with cosmological time. One method is to measure the CCSN rate. Measuring this rate directly beyond local distances has been limited until recent decades. Another method to probe the core-collapse rate is by directly relating $R_{\rm CC}$ to the star formation rate per volume (or SFRD). Since massive stars undergo core collapse rather quickly compared to cosmological timescales, $R_{\rm CC}\propto~{\rm SFRD}$. Well-known compilations of SFRD measurements were carried out in Ref.~\cite{Hopkins:2006bw} in 2006 and again in Ref.~\cite{Madau:2014bja} in 2014. Typically, studies fit these data to functional forms that are integrated over in order to calculate the DSNB rate. But after an additional $\sim10\,{\rm years}$, we have compiled a new, larger list of measurements with two goals: (\textit{i}) to further understand the nuances of cosmic evolution beyond any functional forms, and (\textit{ii}) how the uncertainties have evolved which would lead to more precise DSNB predictions. In Sec.~\ref{subsec:calculations}, this will allow us to better quantify the uncertainty on the DSNB rates from SFRD measurements, alongside neutrino emission and the failed black hole channel.

\begin{table*}
\caption{\label{tab:sfrdtable}The first five rows in the SFRD measurement compilation; the remaining entries can be found at \url{https://github.com/nekanger/Ekanger2023}. For each reference, we show the redshift range, SFRD value and quoted error, indicator, AGN corrections (if any), extinction corrections (if any), and IMF. For each, we also provide the cosmological assumptions, any metallicity corrections, and further details in the larger table. All data reported here are the original, uncalibrated measurements before correcting for IMF and cosmological assumptions. We reference the following works in our full table: Refs.~\cite{Takahashi:2007ia,Morioka:2008ra,Shioya:2007kx,Villar:2007bx,Dunne:2008tk,Shim_2009,Smolcic:2008br,sobral2009,Zhu:2008ee,dale2010,rodighiero2010,vaccari2010,westra2010,Bayliss2011,bothwell2011,karim2011,ly2011,robotham2011,cucciati2012,sobral2012,Burgarella:2013jaa,Ciardullo:2013yka,Gunawardhana:2013fha,Gruppioni:2013jna,Magnelli:2013bya,Sobral:2013fga,Planck:2013qqi,Gruppioni:2015bla,khostovan2015,marchetti2016,Sobral:2015tla,davies2016,guijarro2016,delgado2016,rr2016,vansistine2016,Driver:2017qne,novak2017,audcentross2018,coughlin2018,Fermi-LAT:2018lqt,fernandez2018,sanchez2019,upjohn_brown_hopkins_bonne_2019,bellstedt2020,Gruppioni:2020vue,khostovan2020,vilellarojo2021,Yan:2022ctq,cochrane2023}.}
\begin{ruledtabular}
\begin{tabular}{lcccccl}
\textrm{Redshift}&\textrm{SFRD}&Indicator&AGN&Extinction&IMF&\textrm{Reference}\\
(z)&(${\rm M_{\odot}~yr^{-1}~Mpc^{-3}}$)&&(Y/N)&(Y/N)&\\
\hline
    \lbrack 1.17, 1.2\rbrack&$0.320_{-0.04}^{+0.06}$&O[II]&Y&Y&Salpeter&Takahashi \textit{et al}. 2007 \cite{Takahashi:2007ia}\\
    0.24&$0.035_{-0.014}^{+0.024}$&H$\alpha$&Y&Y&Salpeter&Morioka \textit{et al}. 2008 \cite{Morioka:2008ra}\\
    \lbrack 0.233, 0.251\rbrack&$0.018_{-0.004}^{+0.007}$&H$\alpha$&Y&Y&Salpeter&Shioya \textit{et al}. 2008 \cite{Shioya:2007kx}\\
    0.84&$0.17_{-0.03}^{+0.03}$&H$\alpha$&N&Y&Salpeter&Villar \textit{et al}. 2008 \cite{Villar:2007bx}\\
    $0.12^{+0.08}_{-0.12}$&$0.009_{-0.006}^{+0.006}$&Radio&N&N&Salpeter&Dunne \textit{et al}. 2009 \cite{Dunne:2008tk}\\
\end{tabular}
\end{ruledtabular}
\end{table*}

Observations commonly use UV, IR, and H$\alpha$ line emission as indicators of star formation. However for completeness, we compiled an exhaustive list of SFRD measurements (roughly post-2006, following Ref.~\cite{Madau:2014bja}, and extending until 2023), including also additional indicators such as gamma ray, radio, and other combined methods. We excluded studies for measurements of redshift $z>2$, however, as the neutrinos from high redshifts do not contribute significantly to the detectable DSNB (see Sec.~\ref{subsec:calculations}). We also excluded any studies that did not report an error on the SFRD values. Lastly, we did not include studies that focus on atypical galaxies, including protoclusters, active galactic nuclei (AGNs), or starburst galaxies, or studies that do not measure the total star formation rate of galaxies. Some studies make different assumptions for the limiting star formation rate that is integrated down to when calculating SFRD, so in those cases we choose the smallest reported. While we did not attempt to correct for the different assumptions for the limiting star formation rate, it should be noted that they typically have an effect on the order $\sim10\%$ or less (see, e.g., Ref.~\cite{Westra_2009}).

We systematically correct for two factors in the SFRD data; initial mass function (IMF) and cosmological assumptions. To convert a measured luminosity to a star formation rate, studies have to assume an IMF (commonly, e.g., the Salpeter \cite{Salpeter:1955it}, Chabrier \cite{Chabrier:2003ki}, and Baldry-Glazebrook \cite{Baldry:2003xi} IMFs). Throughout this work, we assume a Chabrier IMF. The choice reflects its wide use and also because it does not have low-mass issues found in the Salpeter IMF \cite{Salpeter:1955it}. In order to convert the results with a Salpeter IMF to a Chabrier IMF, we multiply by 0.63 \cite{Madau:2014bja}. To convert the results with a Baldry-Glazebrook IMF to a Chabrier, we divide by 0.55 which brings it in line with a Salpeter IMF, then multiply by 0.63 \cite{Horiuchi:2008jz}. Although the choice of IMF scales the SFRD measurements, the dependence on the IMF is approximately canceled out when converting to a core-collapse rate, so the final DSNB event rate does not depend strongly on the choice of IMF. However, it is important to be consistent in the IMF used when compiling data. For cosmology we assume $H_0=70\, {\rm km~s^{-1}~Mpc^{-1}},\Omega_M=0.3$, and$~\Omega_{\Lambda}=0.7$, and correct measurements with hubble constants, although this makes a smaller effect on the final results. We show the first five rows of our compilation in Table~\ref{tab:sfrdtable}, with the columns of redshift, SFRD measurement, indicator, whether or not an AGN correction is included, whether or not dust extinction is accounted for, and the associated reference. The full table, with additional details can be found at \url{https://github.com/nekanger/Ekanger2023}. Note that the data provided there is uncorrected for IMF and cosmological assumptions.

\begin{figure*}
\includegraphics[width=\linewidth]{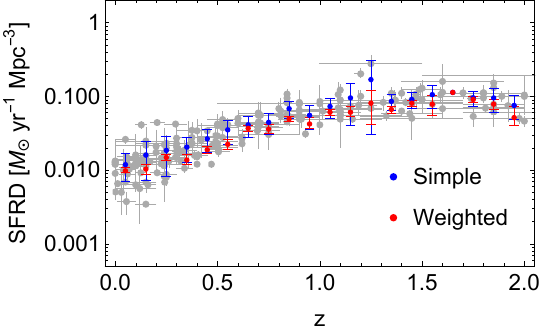}
\caption{SFRD measurements (in gray, from Fig.~\ref{fig:sfrdindicators}) alongside the average values in each redshift bin. We show two averaging schemes: the average values and their standard deviations using simple average (blue) and the inverse-variance-weighted average and their errors (in red). In general, the weighted averages are systematically slightly lower than the simple averages. Also, since in the weighted average case we divide by the number of measurements, $N$, (compared to $N^{1/2}$ in the simple case), the error bars are also generally smaller for the weighted scheme.}
\label{fig:avgsfrd}
\end{figure*}

In Fig.~\ref{fig:sfrdindicators}, we show the results of this latest compilation, a total of 224 SFRD measurements, where the indicator of star formation is given by color and marker style. We separate gamma rays, UV, IR, radio, and H$\alpha$; `Other UVOIR' refers to studies that combine multiple indicators within UV, optical, and IR bands and `Other emission' refers to other emission line measurements, like O[II], O[III], and H$\beta$. Overall, across indicator, IMF, extinction, AGN contamination, and other assumptions, there is fairly good agreement and measurements are generally within error bars of others. The measurements differ by a factor of few and this tends to increase with higher redshifts. There is a hint of systematic issues with other emission line measurements, which may need to be recalibrated with other indicators. Some studies we reference in this compilation include measurements of SFRD past $z>2$, although we do not include them in the list. At these higher redshift cases, the uncertainty is much greater, but, fortunately, the contribution to the DSNB event rate in the detectable range of current and upcoming detectors is negligible beyond $z>2$.

In order to calculate the DSNB in Sec.~\ref{sec:detection}, we bin the SFRD data and average the data within those bins. We choose to bin the data by redshift bins of width $\Delta z=0.1$. This results in around $\sim10$ data points in each bin. Many studies provide a range of redshift values for their SFRD measurement so we take the redshift value in the middle of these ranges (i.e., the average between the upper and lower limits). For the others, we take the quoted value of the redshift. Within each redshift bin, we take two approaches to calculate the average SFRD: a simple average and an inverse-variance-weighted average.

\begin{figure*}
\includegraphics[width=\linewidth]{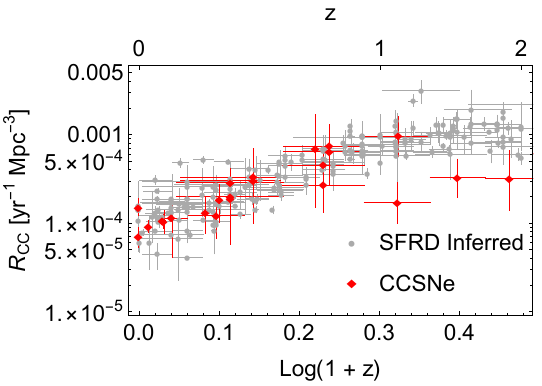}
\caption{Inferred CCSNe rates from recent SFRD measurements (in gray, from Fig.~\ref{fig:sfrdindicators}) compared to direct CCSNe rate measurements (Refs.~\cite{Dahlen:2004km,Cappellaro:2004ti,Botticella:2007er,bazin2009,li2011,graur2011,mattila2012,melinder2012,dahlen2012,Taylor:2014rlo,Graur:2014bua,Cappellaro:2015qia,Strolger:2015kra}, in red). These agree well at low redshifts, but differ between $1<z<2$. However, the number of surveys at these higher redshifts is low (see Ref.~\cite{Strolger:2015kra}) and the contribution to the DSNB drops off significantly above $z>1$. Note the change in the horizontal axis scale to Log(1 + z) from the axes in Figs.~\ref{fig:sfrdindicators} and \ref{fig:avgsfrd}. This was done to highlight the $z<1$ region [$\sim$Log(1 + z) $\approx0.3$].}
\label{fig:ccsnesfrd}
\end{figure*}

When taking a simple average, we take each measurement at face value and use averages in redshift bins. To get an estimate of the uncertainty with this method, we take the standard deviation of each of the measurements. We also adopt a weighted average scheme. For this, we choose the inverse variance for weights. That is, we take the inverse-error-squared of each measurement as weights. While each measurement has errors bars, the upper and lower error bars are not always equal, so for these unequal cases we take the average of upper and lower errors. To estimate the error in each redshift bin, however, we do account for the upper and lower limits separately. To estimate the total error for each bin, we sum each measurement error in quadrature and divide by the number of measurements in each bin. 

In Fig.~\ref{fig:avgsfrd}, we show the average value of the SFRD in each redshift bin using the simple (in blue) and weighted (in red) average methods, with the original measurements in gray. The weighted average estimates are systematically slightly lower than the simple average estimates. The error bars in the `Weighted' case are also smaller compared to the `Simple' case because those errors are proportional to the number of measurements, $N$ (whereas they are proportional to $N^{1/2}$ in the simple case). We focus on the simple averaging case in this study, but discuss the impact of the weighted average in Appendix~\ref{sec:weightedavg}.

\subsection{\label{subsec:directcc}Direct CCSNe measurements}

Another way to measure the rate of core collapse is to utlize CCSN observations. However, these are blind to failed supernovae - massive stars that undergo core collapse and do not successfully power a luminous explosion. Further, only recently have surveys been able to measure sufficient numbers of CCSNe at large distances to probe this measure as a function of redshift. Here, we describe the compilation of studies that measure this quantity out to redshift $z\sim2$, which is the necessary redshift range for the detectable DSNB rate.

We convert our SFRD measurements into rates of core collapse, $R_{\rm CC}$, by
\begin{equation}\label{ccrate}
    R_{\rm CC}=\Dot{\rho_*}(z)\frac{\int_{8M_{\odot}}^{100M_{\odot}}\psi(M)dM}{\int_{0.1M_{\odot}}^{100M_{\odot}}M\psi(M)dM},
\end{equation}
where $\Dot{\rho_*}(z)$ is the star formation rate density. In this conversion, we have assumed that stars with masses above $8\,M_{\odot}$ and up to $100\,M_{\odot}$ undergo core collapse. If all of these core collapses produce CCSNe, then we can directly compare with the observed CCSN rates. If an appreciable fraction of core collapse produce failed CCSNe, then we expect the observed CCSN rate to fall short. 

We compile a list of studies complete from the years $\gtrsim2000$ (Refs.~\cite{Dahlen:2004km,Cappellaro:2004ti,Botticella:2007er,bazin2009,li2011,graur2011,mattila2012,melinder2012,dahlen2012,Taylor:2014rlo,Graur:2014bua,Cappellaro:2015qia,Strolger:2015kra}) until today. In Fig.~\ref{fig:ccsnesfrd}, we show how the SFRD-inferred rates compare to our catalog of directly measured CCSN rates out to redshift $z=2$ (or Log($1+z$) $\sim0.5$, note the change in axes from Figs.~\ref{fig:sfrdindicators} and \ref{fig:avgsfrd}). These direct measurements agree well within $z\lesssim1$ (in contrast with older data \cite{Horiuchi:2011zz} but in agreement with more recent comparisons, see, e.g., \cite{Dahlen:2012cm,Madau:2014bja,Mathews:2014qba}), but the direct measurements from Ref.~\cite{Strolger:2015kra} between $1<z<2$ appear to be systematically lower than those inferred from SFRD measurements. Fortunately, the DSNB event rates are not as affected by this redshift regime, so the difference is not as consequential (see Sec.~\ref{subsec:calculations}), but further high-redshift supernova studies will be necessary to resolve these conflicting measurements.

Although there is good agreement between direct CCSN rate measurements and the SFRD inferred $R_{\rm CC}$ measurements, we choose to use the latter alone for the calculation of the DSNB. The sample of direct measurements is relatively small and many have large error bars. Further, combining the two measurements could introduce systematic issues.

\section{\label{sec:neutrinoemission}Neutrino Emission from Supernovae}

\subsection{\label{subsec:successful}Successful supernova neutrino emission}

The DSNB can be roughly broken into two source classes: successful supernovae where the shock is revived, and failed supernovae where this shock is not revived. We first discuss how we model the neutrino emission from the former.

To model the neutrino emission from successful supernovae, we use the suite of two-dimensional axisymmetric simulations from Ref.~\cite{Nagakura:2021lma}. This simulation set is particularly useful because they have been carried out for $\sim4\,{\rm s}$ postbounce, and understanding the neutrino emission over $\mathcal{O}(10)\,{\rm s}$ is essential for accurately modeling the DSNB rate. The simulation set includes 15 progenitors of masses ranging from $13\,M_{\odot}$ to $26.99\,M_{\odot}$ (the $12\,M_{\odot}$ and $15\,M_{\odot}$ models, however, do not explode so we do not count them as successful supernovae). The SFHo equation of state (EOS) \cite{Steiner:2012rk} is employed in these simulations. We supplement the simulations with additional lower mass, 2D simulations for progenitors of $9,~10,$ and $11\,M_{\odot}$ (also $\sim$ seconds long and provided by D. Vartanyan, private communication) and the $8.8\,M_{\odot}$ electron-capture SN model from Ref.~\cite{hudepohl2010}. 

Approximately half of the neutrinos emission occurs during the cooling phase, i.e., when the newly formed hot protoneutron star (PNS) cools via neutrino heating (occurring $\gtrsim1\,{\rm s}$ postbounce). Indeed, the DSNB rate can vary up to a factor of a few due to the uncertainty in the PNS cooling phase \cite{Ekanger:2022neg}. Thus, we model the neutrino emission after the available $\sim4\,{\rm s}$ of simulation data as well. To do so, we follow the `Analytic' method of Ref.~\cite{Ekanger:2022neg} based on  analytic solutions. In Ref.~\cite{Suwa:2020nee}, analytic functions to describe the neutrino luminosity and mean energy are derived assuming spherical symmetry and a thermal emission spectrum. These functions depend on physical parameters like PNS mass, radius, and energy released. They also depend on a density correction factor ($g$) and an opacity boosting factor ($\beta$). We allow these factors to vary such that they can be tuned to best match simulation data between the shock revival time and the end of simulation data. We then calculate the total neutrino energy liberated and mean neutrino energy by integrating the simulation data up to $\sim4\,{\rm s}$ (the end of simulation time postbounce) and then integrating the analytic solutions from $\sim4\,{\rm s}$ to $20\,{\rm s}$ so that we have a complete picture of the average neutrino emission over $20\,{\rm s}$. We do this separately for all neutrino flavors: $\nu_e$, $\bar{\nu}_e$, and $\nu_x$ (where $\nu_x$ represents all nonelectron-type flavor neutrinos). 

\begin{figure}
\includegraphics[width=\linewidth]{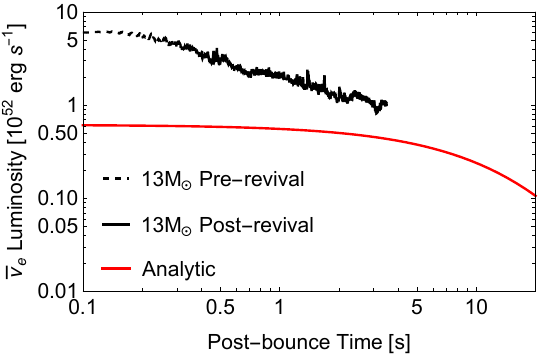}
\includegraphics[width=\linewidth]{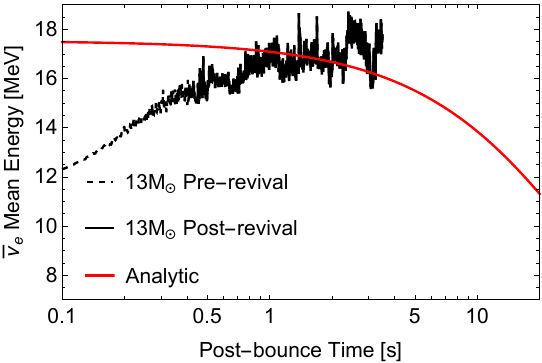}
\caption{Results of the `Analytic' method---tuning analytic supernova light curve solutions to the simulation data---for a $13\,M_{\odot}$ progenitor. In the top panel we show the $\bar{\nu}_e$ luminosity data in black, where data before (after) the time of shock revival is dashed (solid), and the tuned analytic function in red. Similarly, in the bottom panel we show the mean energy data in black and the tuned analytic function in read.}
\label{fig:analyt}
\end{figure}

\begin{figure}
\includegraphics[width=\linewidth]{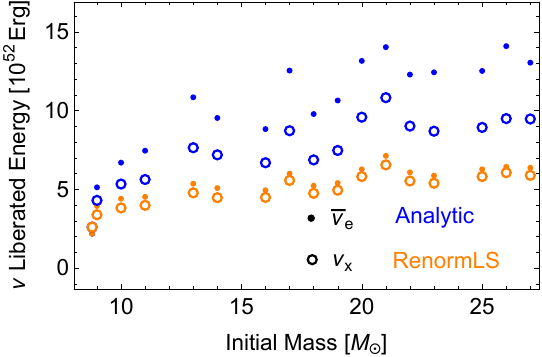}
\includegraphics[width=\linewidth]{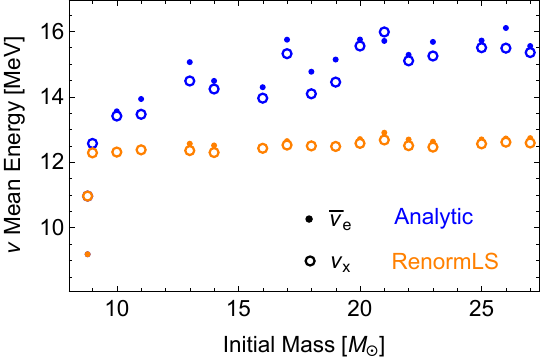}
\caption{Integrated neutrino quantities for each of the progenitors in our adopted simulation set, including the low-mass $8.8\,M_{\odot}$ progenitor (far left point in upper and lower panels). In the top panel, we plot the energy liberated for the $\overline{\nu}_e$ (solid markers) and $\nu_x$ (open markers) flavors as a function of initial mass. In the bottom panel, we plot the mean energies as a function of initial mass. In both panels, the points in blue are showing the values for the `Analyic' method (see Fig.~\ref{fig:analyt}) while orange points show the results for the `RenormLS' method. In general, the `RenormLS' correlation-based method results in lesser liberated energies and lower mean energies, resulting in lower DSNB signal rates. We take the `Analytic' method as the fiducial method to estimate the late-phase neutrino signal.}
\label{fig:integrated}
\end{figure}

In Fig.~\ref{fig:analyt}, we show the result of tuning the analytic solution function to the available data for neutrino luminosity and mean energy, for electron antineutrinos, $\overline{\nu}_e$. In the top panel, we show the simulation data for the neutrino luminosity in black for a $13\,M_{\odot}$ progenitor. We tune the analytic function to the mean energy simulation data only after shock revival time, shown in solid (data before, not used, is shown in dashed). The resulting function is plotted in red (same tuning parameters for both luminosity and mean energy). Note we show the function down to 0.1 seconds but it is invalid in such early phases and are only to be used post shock revival. The analytic function does not fit the luminosity well which may be because of continued fallback accretion onto the PNS, which boosts neutrino luminosity \cite{Akaho:2023alv}. This occurs in multidimensional simulations naturally, so this could explain the discrepancy with the analytic function, which assumes spherical symmetry. The analytic function does fit the mean energy data fairly well, however, especially after shock revival time. We can then extrapolate past the simulation time to infer how the neutrino energy cools over time. These results are in general similar for $\nu_e$ and $\nu_x$ as well.

Although we use the analytic method as our fiducial method to estimate the neutrino emission after simulation end time, we also estimate the late-phase neutrino emission using the `RenormLS' method of Ref.~\cite{Ekanger:2022neg}. This second method uses a two-parameter correlation (final PNS mass and shock revival time), which is based on long-term PNS cooling simulations to estimate the neutrino emission after shock revival time. Although both of these estimations are based on one-dimensional simulations, this is suitable for modeling the $\mathcal{O}(10)\,{\rm s}$ cooling phase of neutrino emission and still respects the early, accretion phase of our axisymmetric simulations (for dimension-dependent results in the early, accretion-dominated phase, see, e.g., Ref.~\cite{Nagakura:2020qhb}).

In Fig.~\ref{fig:integrated}, we show the integrated neutrino quantities for all progenitors using both methods for $\overline{\nu}_e$ (solid markers) and $\nu_x$ (open markers). In the top panel, we show the time-integrated liberated energy and in the bottom panel we show the time-averaged mean energy. For both panels, in blue we show the results for the `Analytic' method and in orange are the results for the `RenormLS' method. The `Analytic' method (see Fig.~\ref{fig:analyt}) predicts systematically higher liberated and mean energies compared to the `RenormLS' method. This `RenormLS' method represents a more conservative scenario where the PNS cooling phase occurs earlier and results in lower neutrino emissions. Although we take the `Analytic' method to be our fiducial case, we discuss the impact of the late-phase treatment in Sec.~\ref{sec:detection}.

\subsection{\label{subsec:failed}Failed supernova neutrino emission}

We now discuss modeling the neutrino emission from failed supernovae. This case may be just as important to model as the successful case, as up to $\gtrsim40$\% of core collapses may fail (see Refs.~\cite{OConnor:2010moj,Gerke_2015,Adams_2017,Kresse:2020nto,Neustadt_2021,Burrows:2023nlq}), but the exact fraction ($f_{\rm BH}$) is not well-known. Further, the criterion for failed supernova may not be as simple as the initial progenitor mass (see alternative criteria such as the compactness parameter \cite{OConnor:2010moj}, the enclosed mass-dimensionless entropy parameters \cite{Ertl:2015rga}, the force explosion condition \cite{Gogilashvili:2021xfe}, and the density jump at the silicon-oxygen layer \cite{Ott:2017kxl,Wang:2022dva,Boccioli:2022ktx}). For these reasons, we adopt as our fiducial value $f_{\rm BH}=23.6$\% \cite{Neustadt_2021} obtained from an observational survey of disappearing massive supergiants \cite{Kochanek:2008mp,2015MNRAS.450.3289G,Adams:2016ffj,Adams:2016hit,Basinger:2020iir}. The survey found two candidates, and our adopted value assumes both are failed supernovae. 

The neutrino emission from failed supernovae is also very uncertain, primarily due to model and EOS dependences. For example, for a $40\,M_{\odot}$ progenitor, the neutrino emission/light curves from fallback accretion can vary depending on the code used, model assumptions, and metallicity (see, e.g., Refs.~\cite{Chan:2017tdg,Ott:2017kxl,Moriya:2019jon,Burrows:2023nlq}, Refs.~\cite{Kuroda:2018gqq,Kuroda:2023mzi} for additional simulations, and the effect this may have on the DSNB in Ref.~\cite{Lunardini:2009ya}). The EOS plays a large role in determining the time to black hole formation, and, thus, the total and average neutrino energy emitted \cite{Nakazato:2021gfi}. We use the $40\,M_{\odot}$ failed supernova neutrino emission data from Ref.~\cite{Walk:2019miz}, which assumes the LS220 EOS \cite{1991NuPhA.535..331L}, for our fiducial case. We also consider the $30\,M_{\odot}$ models with different equations of state (Shen and LS220 at 1/5 solar metallicity) from Ref.~\cite{Nakazato:2021gfi} to get a sense of the uncertainty from different failed supernova models. These Shen and LS220 models give rise to relatively small and very high mean energies, respectively, which represent the extremes of the failed supernova models.

\section{\label{sec:detection}Signal Prediction and Detection Prospects}

\subsection{\label{subsec:calculations}Calculating event rates}

With estimates of the total neutrino energy liberated and mean energy from Secs.~\ref{subsec:successful} and \ref{subsec:failed}, we can calculate the average neutrino emission spectrum. Although we have methods to estimate the mean energy (first moment) from simulations and analytic model, the second moment remains less reliable. Because of this, we assume a pinched Fermi-Dirac spectrum, $f(E)$, with a pinching parameter value of $\alpha=2.3$ to approximate a thermal Fermi-Dirac spectrum, where \cite{Keil:2002in}:
\begin{equation}\label{pinchedfd}
    f(E)=\frac{(1+\alpha)^{1+\alpha}}{\Gamma(1+\alpha)}\frac{E_\nu E^{\alpha}}{(\epsilon_\nu)^{2+\alpha}}\exp\left[-(1+\alpha)\frac{E}{\epsilon_\nu}\right].
\end{equation}
Although the simulations from Ref.~\cite{Nagakura:2021lma} produce neutrino energy spectra, we simplify the analysis with this pinched Fermi-Dirac spectrum as this does not largely change our event rates. Once we have an approximate spectrum for each progenitor, we calculate the IMF-weighted average spectrum, $dN/dE$, where:
\begin{equation}\label{sourcespectrum}
    \frac{dN}{dE}=\sum_i\frac{\int_{\Delta M_i}\psi(M)dM}{\int_{M_0}^{M_f}\psi(M)dM}f_i(E).
\end{equation}
Here, the subscript $i$ represents each (successful) progenitor we consider, $f_i(E)$ is that progenitor's spectrum from Eq.~(\ref{pinchedfd}), $\Delta M_i$ is that progenitor's corresponding mass bin width, $M_0=8\,M_{\odot}$ is our lower integration limit, $M_f=40\,M_{\odot}$ is our upper integration limit for successful supernovae, and $\psi(M)$ is the initial mass function. For our lowest-mass progenitor ($8.8\,M_{\odot}$ electron-capture supernova), we take $\Delta M_i=[8\,M_{\odot},~8.9\,M_{\odot}]$, for the bins of progenitor masses $13\,M_{\odot}$ to $26\,M_{\odot}$ we take $\Delta M_i=[(M_{i-1}+M_i)/2,~(M_i+M_{i+1})/2]$, and for our highest-mass progenitor ($26.99\,M_{\odot}$), we take $\Delta M_i=[26.5\,M_{\odot},~40\,M_{\odot}]$. This gives a total of 17 mass bins. As in Sec.~\ref{sec:rcc}, we choose the Chabrier IMF for $\psi(M)$.

Next, we incorporate our updated star formation rate density measurements to calculate the DSNB flux and event rate at Earth. The flux is given by
\begin{equation}\label{flux}
    \frac{d\phi}{dE}=c\int R_{\rm CC}(z)\frac{dN}{dE'}(1+z)\left|\frac{dt}{dz}\right|dz,
\end{equation}
where $E'=E(1+z)$ and $|dz/dt|=H_0(1+z)[\Omega_m(1+z)^3+\Omega_{\Lambda}]^{1/2}$. In Sec.~\ref{subsec:sfrd}, we compiled a list of SFRD measurements that are corrected to the same Chabrier IMF. Here, we bin the measurements in redshift and, rather than an integral, sum up the redshift bins of width $\Delta z=0.1$. 

To account for neutrino emission from both successful and failed supernovae, we compute the total flux as the sum,
\begin{equation}
\frac{d\phi}{dE}\Big|_{\rm tot}=
(1-f_{\rm BH})\frac{d\phi}{dE}\Big|_{\rm s}+f_{\rm BH}\frac{d\phi}{dE}\Big|_{\rm f},
\end{equation}
where subscripts $s$ and $f$ indicate successful and failed supernova, respectively. We also consider the Mikheyev-Smirnov-Wolfenstein (MSW) effect for neutrino oscillations,
\begin{equation}
\begin{split}
\frac{d\phi}{dE}^{\rm obs}_{\nu_e}\approx
\begin{cases}
    \frac{d\phi}{dE}_{\nu_x}~({\rm NO}), \\
    \frac{d\phi}{dE}_{\nu_e}\sin^2\theta_{12}+\frac{d\phi}{dE}_{\nu_x}\cos^2\theta_{12}~({\rm IO}),\\
\end{cases}
\end{split}
\end{equation}

\begin{equation}
\begin{split}
\frac{d\phi}{dE}^{\rm obs}_{\bar{\nu}_e}\approx
\begin{cases}
    \frac{d\phi}{dE}_{\bar{\nu}_e}\cos^2\theta_{12}+\frac{d\phi}{dE}_{\nu_x}\sin^2\theta_{12}~({\rm NO}),\\
    \frac{d\phi}{dE}_{\nu_x}~({\rm IO}),
\end{cases}
\end{split}
\end{equation}
where NO and IO are the normal and inverted mass orderings, respectively, $\cos^2\theta_{12}\approx0.7$ and $\sin^2\theta_{12}\approx0.3$ (assuming $\sin^2\theta_{13}\ll1$, see e.g., Ref.~\cite{2018moller}).

\begin{figure}
\includegraphics[width=\linewidth]{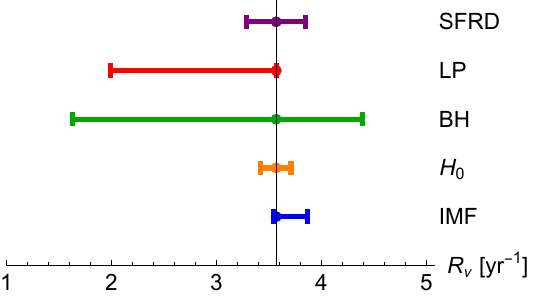}
\caption{Estimated errors of DSNB event rates for normal ordering at SK-Gd from SFRD measurements (`SFRD'), late-phase treatment (`LP', Analytic, or RenormLS), failed supernova modeling (`BH', see Fig.~\ref{fig:bhfraction}), $H_0$, and IMF assumption (`IMF', Chabrier, Salpeter A, or Baldry-Glazebrook). Quantitative values are given in Table~\ref{tab:eventserrors}.}
\label{fig:errorplot}
\end{figure}

\renewcommand{\arraystretch}{1.5}
\begin{table}[b]
\caption{\label{tab:eventserrors}Integrated DSNB rate and flux ($\phi$) with errors, in the SK-Gd energy range ($9.3 < E_{\nu} < 31.3\,{\rm MeV}$, where $E_{\nu}$ is the neutrino energy).
Values shown for both normal and inverted ordering. SFRD error calculated directly from measurements. Error for the ``late phase'' (LP) calculated by assuming the RenormLS method (whereas the fiducial value is calculated assuming the Analytic method). BH fraction error naturally takes into account the late phase method chosen, failed supernova model, different equations of state, and $f_{\rm BH}$. We take `BH' error as the maximum and minimum values within the blue-shaded trapezoid in Fig.~\ref{fig:bhfraction}. At SK-Gd, we expect around 10.6 background events per year \cite{Super-Kamiokande:2023xup} with 0.01\% Gd, but can be reduced with convolutional neural networks \cite{Maksimovic:2021dmz}.}
\begin{ruledtabular}
\begin{tabular}{cccccc}
&Ordering&Fiducial&SFRD err&LP err&BH err\\
\colrule
$R_{\nu}$&\textrm{NO}&3.57&$^{+0.28}_{-0.28}$&$^{+0.00}_{-1.58}$&$^{+0.82}_{-1.94}$\\
{[${\rm yr^{-1}}$]}&\textrm{IO}&2.85&$^{+0.28}_{-0.28}$&$^{+0.00}_{-1.07}$&$^{+0.17}_{-1.30}$\\
\hline
$\phi$&\textrm{NO}&5.10&$^{+0.40}_{-0.40}$&$^{+0.00}_{-2.04}$&$^{+0.50}_{-2.70}$\\
{[${\rm cm^{-2}~s^{-1}}$]}&\textrm{IO}&4.10&$^{+0.40}_{-0.40}$&$^{+0.00}_{-1.36}$&$^{+0.20}_{-1.92}$\\
\end{tabular}
\end{ruledtabular}
\end{table}

\begin{table}[b]
\caption{\label{tab:ratesother}Table quantifying the fiducial mass, detection energy range, corresponding DSNB signal rate ($R_{\nu}$) and flux ($\phi$) for normal (and inverted) ordering for JUNO (see Ref.~\cite{Li:2022myd}), HK and HK-Gd (see Ref.~\cite{Sawatzki:2020mpb}), and DUNE (see Ref.~\cite{Sawatzki:2020mpb}).}
\begin{ruledtabular}
\begin{tabular}{cccccc}
&Ordering&Mass&$E_{\nu}$&$R_{\nu}$&$\phi$\\
&&[kton]&[MeV]&[${\rm yr^{-1}}$]&[${\rm cm^{-2}~s^{-1}}$]\\
\colrule
JUNO&NO (IO)&17&12-30&2.07 (1.65)&2.88 (2.29)\\
\hline
HK&NO (IO)&187&20-30&7.70 (6.10)&0.50 (0.39)\\
HK-Gd&NO (IO)&187&10-30&27.60 (22.00)&4.39 (3.50)\\
\hline
DUNE&NO (IO)&40&16-40&5.70 (5.30)&1.02 (0.97)\\
\end{tabular}
\end{ruledtabular}
\end{table}

Finally, we multiply the flux by the cross section of the corresponding experiment ($\sigma_{\nu}$) and the number of target protons, before integrating over the detection energy window, to obtain the DSNB event rate
\begin{equation}\label{eventrate}
    R_{\nu}=N_t\int dE\frac{d\phi(E_{\nu})}{dE}\sigma_{\nu}(E_{\nu}).
\end{equation}
For SK doped with gadolinium (SK-Gd), $N_t=1.5\times10^{33}$ is the number of target protons, we take $\sigma_{\nu}$ as the IBD cross section (see Refs.~\cite{Vogel:1999zy,Strumia:2003zx}), and we adopt the detection energy range for SK-Gd from $9.3\,{\rm MeV}$ to $31.3\,{\rm MeV}$ \cite{Super-Kamiokande:2023xup}. 

In Table~\ref{tab:eventserrors}, we present the DSNB flux and yearly event rates at SK-Gd for the normal and inverted mass orderings (see Appendix~\ref{sec:unoscillated} for unoscillated results). The column `SFRD err' displays the statistical error from our SFRD compilation. To do this for the simple- and weighted-average cases, we sum the errors in each redshift bin in quadrature (see Sec.~\ref{subsec:sfrd} for how the errors are calculated in each redshift bin). The `LP err' column displays the error due to differing estimates of the late-time neutrino emission. In our modeling, this is estimated by comparing the analytic late-time model to the `RenormLS' method. Finally, the `BH err' column displays the error due to failed supernova. More specifically, we use the $1\sigma$ upper and lower $f_{\rm BH}$ bounds from Ref.~\cite{Neustadt_2021}. We also represent these results and more in Fig.~\ref{fig:errorplot}. The vertical line shows our fiducial estimate for the yearly DSNB signal rate at SK-Gd and the different horizontal lines show the estimated errors in Table~\ref{tab:eventserrors} along with the error from varying $H_0$ and the IMF. To calculate the $H_0$ error we take the values of $67.4\,{\rm km~s^{-1}~Mpc^{-1}}$ \cite{Planck:2018vyg} and $73\,{\rm km~s^{-1}~Mpc^{-1}}$ \cite{Riess:2021jrx} and scale the SFRD linearly with $H_0$. To calculate the IMF error, we take Baldry-Glazebrook \cite{Baldry:2003xi} and Salpeter A IMFs (see Ref.~\cite{Hopkins:2006bw}). Both of these are shown in addition to the other errors to highlight that these errors contribute to a $<10\%$ DSNB uncertainty.

\begin{figure}
\includegraphics[width=\linewidth]{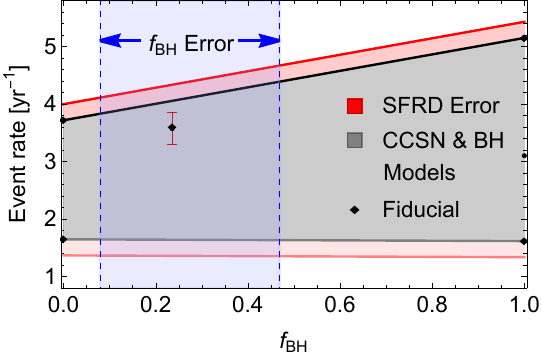}
\caption{DSNB event rate at SK-Gd as a function of failed supernova (or black hole) fraction, $f_{\rm BH}$. We show the fiducial value assuming $f_{\rm BH}=0.236$ with the black diamond point, assuming the BH model from Ref.~\cite{Walk:2019miz}. The other five points at $f_{\rm BH}=0$ and $f_{\rm BH}=1$ represent the extreme cases: at $f_{\rm BH}=0$, the upper point is given by assuming the `Analytic' method for the late phase and the lower point assumes the `RenormLS' method, while at $f_{\rm BH}=1$, the upper point assumes the BH-forming model with Shen EOS from Ref.~\cite{Nakazato:2012qf} and the lower point assumes the BH-forming model with the LS220 EOS from Ref.~\cite{Nakazato:2012qf} (both are $30\,M_{\odot}$ progenitors). The middle point at $f_{\rm BH}=1$ represents the value assuming the $40\,M_{\odot}$ BH-forming model with LS220 EOS from Ref.~\cite{Walk:2019miz}. The shaded gray region in between reflects the combination of late phase treatment and $f_{\rm BH}$. In red, we show the expected error from our collected SFRD measurements. In blue, we show the range of $f_{\rm BH}$ from 0.079 to 0.469 which is the $1\sigma$ error from Ref.~\cite{Neustadt_2021}. These results assume NO and the simple average method for the SFRD data.}
\label{fig:bhfraction}
\end{figure}

In Fig.~\ref{fig:bhfraction}, we show in more detail the failed supernova contribution to the DSNB event rate as a function of the failed fraction. The diamond point in the center shows our fiducial value shown also in Table~\ref{tab:eventserrors}, where the error corresponds to that from the `SFRD error' and $f_{\rm BH}=23.6$\% is adopted from Ref.~\cite{Neustadt_2021} (the blue region denotes the $1\sigma$ uncertainty bounds on $f_{\rm BH}$ from the same study, from $7.9\%$ to $46.9\%$). However, the failed supernova contribution also depends strongly on the failed supernova model and EOS. On the far right, i.e., for a pure failed fraction, we plot the predictions from the $30\,M_{\odot}$ model with LS220 EOS from Ref.~\cite{Nakazato:2012qf}  (highest point) as well as the same $30\,M_{\odot}$ model but with Shen EOS \cite{Shen:1998gq} (lowest point). Note, the model we choose for our fiducial model (diamond point) is the $40\,M_{\odot}$, 3D, LS220 EOS model from Ref.~\cite{Walk:2019miz}, which lies between these. On the far left, i.e., for no failed supernovae, we plot the predictions using two differing late-time estimates (analytical versus RenormLS). All predictions therefore lie within the trapezoid defined by these extremities. Depending on the true failed supernova model as well as successful supernova model, the neutrino mean energies in particular vary significantly, causing the DSNB event rate to either decrease or increase with increasing $f_{\rm BH}$. For these reasons, the BH error in Fig.~\ref{fig:errorplot} is estimated as the largest uncertainty source at around $\sim50$\%. 

\begin{figure}
\includegraphics[width=\linewidth]{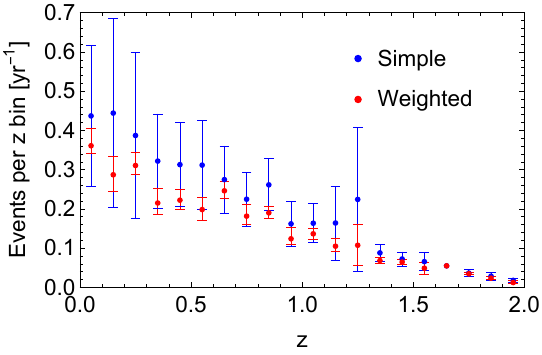}
\caption{Event rates per year for each redshift bin, assuming the SK-Gd energy window. In blue is the calculation for the simple averaging method while red is the calculation for the weighted averaging method. Most of the DSNB signal comes from redshifts $z<1$ and becomes negligible above $z=2$.}
\label{fig:redshiftcontribution}
\end{figure}

We also show the contribution to the DSNB rate per redshift bin in Fig.~\ref{fig:redshiftcontribution}. The bins are of width $\Delta z=0.1$ and the values are found by either taking a simple arithmetic mean of the SFRD data in each bin (blue) or a weighted mean (red, weighted by the inverse-error-squared) in each bin. In either case, the biggest contribution to the detectable DSNB energy window comes from $z\lesssim1$ and diminishes to negligible by $z\approx2$, validating our redshift cutoff for SFRD data around $z\approx2$.

Lastly, in Table~\ref{tab:ratesother}, we also show the expected DSNB signal rate per year, $R_{\nu}$, for JUNO, HK, HK-Gd, and DUNE experiments. The backgrounds for these experiments are not well characterized yet, but we can predict the signal rate for the given masses and energy ranges. For DUNE, we assume the $\nu_e-{\rm Ar}$ cross section from Ref.~\cite{Cocco:2004ac}. 

\subsection{\label{subsec:significantdetection}Detection prospects}

Here, we incorporate estimates of the backgrounds for the SK-Gd and JUNO detectors in order to forecast how long it would take to significantly detect the DSNB. Because we do not know the backgrounds well for HK and DUNE, we leave them out of the following discussion. Following Ref.~\cite{Li:2022myd}, we use a figure of merit significance of $\sim S/\sqrt{S+B}$ where $S$ is the signal event rate and $B$ is the background event rate. For the first phase of SK-Gd with 0.01\% Gd, we do a binned analysis since we have available background data \cite{Super-Kamiokande:2023xup}. For the second phase of SK-Gd with 0.03\% Gd, we follow the techniques from Ref.~\cite{Li:2022myd} and conservatively assume that the background is reduced to a $S:B$ ratio of 2:1 following an implementation of convolutional neural networks into the SK-Gd analysis. For JUNO we do a rate-only analysis, i.e. an analysis with only one energy bin, and again assume a $S:B$ ratio of 2:1. 

For the signal, we take the rates from Sec.~\ref{subsec:calculations}, integrated over the appropriate energy bin range, but include a treatment of the detection efficiency. For SK-Gd, from August 2020 to June 2022 the Gd concentration is  $0.01\%$, and we adopt an average total signal efficiency of $\sim30\%$ (see Fig.~1 of Ref.~\cite{Super-Kamiokande:2023xup}). During this phase, we adopt the expected background rates from Table 1 of Ref.~\cite{Super-Kamiokande:2023xup}, which was around $\sim16$ in total over 552 days between reconstructed energies $7.5\,{\rm MeV}$ to $29.5\,{\rm MeV}$ ($9.3\,{\rm MeV}$ to $31.3\,{\rm MeV}$ in neutrino energy). Recently in the second phase of SK-Gd, the neutron-tagging efficiency has been shown to be around $\sim60\%$ with $0.03\%$ Gd concentration \cite{Nakanishi:2023vf}. The average signal efficiency before neutron-tagging in Ref.~\cite{Super-Kamiokande:2023xup} is $\sim80\%$, so the new average signal efficiency after neutron-tagging is around $80\%\times60\%\sim50\%$. For JUNO we follow Refs.~\cite{JUNO:2015zny,Li:2022myd} and integrate our DSNB signal over the energy range $12\,{\rm MeV}$ to $30\,{\rm MeV}$ with a signal efficiency of $50\%$, and a detector volume of $17\,{\rm kton}$. This results in a signal (background) rate of around $\sim1\,{\rm yr^{-1}}$ ($\sim0.5\,{\rm yr^{-1}}$).

\begin{figure}
\includegraphics[width=\linewidth]{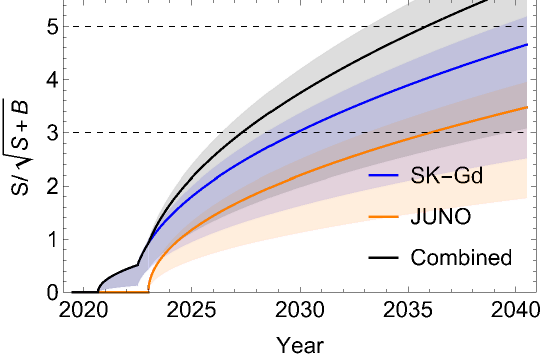}
\caption{A simplified statistical estimate of the significance by which the DSNB can be detected as a function of year for SK-Gd (blue), JUNO (orange), and a combined analysis (black). Based on recent background rate data, a combined SK-Gd and JUNO analysis should be able to detect the DSNB to $3\sigma$ significance by $\sim2030$ with our fiducial calculation and will be complemented by other experiments over the next two decades. The shaded regions show the significance within our SFRD, late phase, and BH uncertainties.}
\label{fig:significanceplot}
\end{figure}

In Fig.~\ref{fig:significanceplot}, we show how the significance metric changes as a function of exposure time for SK-Gd, JUNO, and a combined analysis. This shows that we expect the DSNB to be detected at a level of $3\sigma$ by $\sim2030$ with SK-Gd and a combined SK-Gd/JUNO detection at a level of $5\sigma$ by $\sim2035$. This reflects that more Gd has been dissolved and is increasing the overall efficiency and, thus, the detection significance. This does rely on the ability of convolutional neural networks to help reduce backgrounds, but forecasts suggest this could ultimately result in a $S:B$ ratio of 4:1 \cite{Maksimovic:2021dmz} which would be even better than the 2:1 ratio assumed in Fig.~\ref{fig:significanceplot}. The solid lines show how the significance changes with time for our fiducial signal while the shaded regions reflect our uncertainty from SFRD, late phase, and BH errors (given in Table~\ref{tab:eventserrors} for SK-Gd). With machine learning techniques and more data on SK-Gd and JUNO backgrounds, spectral analyses will be possible and will improve detection significance compared to this more simple rate-only analysis.

We do not make estimates for the detection prospects at HK or HK-Gd because of several quantities to be determined, including backgrounds rates, efficiencies reached, and whether gadolinium will be added. We do know, however, that the signal should increase dramatically with a volume increase to $187\,{\rm kton}$ for HK, but significance ultimately also depends on the backgrounds (see Ref.~\cite{Hyper-Kamiokande:2018ofw} for some predictions on the signal efficiency and detectable energy range). The introduction of other detection channels, for example with electron neutrino flavor scattering at DUNE \cite{DUNE:2015lol}, would improve our understanding of the DSNB models as well. We also do not include DUNE in our analysis here because the background rates and $\nu_e-{\rm Ar}$ cross section are uncertain (see Refs.~\cite{Cocco:2004ac,2018moller,Sawatzki:2020mpb} for more comparisons). Joint analyses between experiments should further improve the DSNB detection significance and understanding of the spectrum.

\section{\label{sec:discussion}Discussion and Conclusions}

In this study, we combined data products from the recent star formation rate measurements and CCSNe simulations to make accurate DSNB flux and event rate predictions. We also quantify the uncertainty in primarily three ingredients: error from rate of core collapse, error from the neutrino emission of typical CCSNe, and error due to the poorly constrained failed CCSN population. These uncertainties will be somewhat degenerate with each other when measuring the event rates at current and next-generation detectors, so understanding their error is critical (however, information such as energy spectra can help break the degeneracy). 

Estimating the core-collapse rate $R_{\rm CC}$ is still more precise by using the SFRD rather than the CCSN directly. Over many decades, several different indicators have been used to measure this quantity and largely agree with each other, within error bars. Improving dust corrections and calibrations between indicators should improve error bars with time. While we collected SFRD measurements systematically, we omitted some measurements, for example those without published errors, those focusing on subsets of galaxies, or those with uncertain assumptions. A special case concerns the SFRD values from Ref.~\cite{bellstedt2023resolving}, where a large number of measurements are given. With so many measurements from one study per redshift bin, this would drive down the error bars to very small values, which may not accurately reflect the true uncertainty. A naive comparison, though, shows that our compilation agree very well with the data from Ref.~\cite{bellstedt2023resolving}, motivating its inclusion in the overall compilation, but excluded from the error analysis.

In parallel, measuring core-collapse rate $R_{\rm CC}$ with future optical instruments will shed light on this quantity directly. The Legacy Survey of Space and Time (LSST) of the Vera C. Rubin Observatory, in particular, may be able to observe possibly thousands of CCSNe each night, up to and exceeding $z\sim1$ \cite{Lien:2009db,Bianco:2014mna,LSST:2008ijt,Heston:2023gbx}. LSST, with other surveys, will give better estimates of $R_{\rm CC}$ and may also shed light on the fraction of supernovae that fail, $f_{\rm BH}$.

In this work, we use the neutrino emission from a suite of of 2D simulations performed until several seconds \cite{Nagakura:2021lma}. These are particularly useful because they probe the neutrino emission to longer timescales around $\sim5\,{\rm s}$. Long term, extensive datasets, like the data found in Ref.~\cite{Vartanyan:2023zlb}, would provide a larger picture of the variability in neutrino emission from many progenitors. Although this large data set does not have the NS radius and shock radius evolution we would need to estimate the late-phase emission, it is useful for understanding which progenitors succeed in exploding and the neutrino emission from them is very similar to the data we use in our study for the same progenitors. Additionally, 3D simulations like those in Ref.~\cite{Wang:2023vkk} and Ref.~\cite{2021bollig} that are carried out to $\sim{\rm few}\,{\rm second}$ may give the most accurate picture of the dynamics and neutrino emission during a CCSNe. The angle averaged neutrino emission, however, is very similar between the 2D simulations we use and the 3D simulations of Ref.~\cite{Wang:2023vkk}, so they appear suitable for this study.

We consider that neutrinos oscillate due to the MSW effect. While this may manifest on average, there are some associated uncertainties. First, the oscillation outcomes depend on the values of neutrino mixing angles. From Ref.~\cite{Capozzi:2017ipn}, the uncertainty in $\sin^2\theta_{12}$ is $<10\%$. This results in an overall uncertainty in $<1\%$ event rate at SK-Gd. Secondly, additional oscillations may result. For example, when supernova neutrinos propagate through the Earth, the $\nu_e$/$\overline{\nu}_e$ flavor ratio may also change the observed spectrum. This deviation may alter the spectrum at a level of $\sim10\%$ \cite{Lunardini:2001pb}. Lastly, collective oscillations may imprint a change in the flux ratios of different flavors. Although the exact oscillation scheme is still under investigation,  Ref.~\cite{Lunardini:2012ne} suggests that collective oscillations affect the signal at the $<10\%$ level. As a concrete example, consider that fast flavor conversions occur deep inside the CCSN core, such that the flavors can be equipartitioned. If flavor equipartition is realized (see Ref.~\cite{Nagakura:2020bbw}, Equations 4 - 7 where $p=\overline{p}=1/3$), the resultant DSNB rate is reduced to $3.19\,{\rm yr^{-1}}$ at SK-Gd for the normal ordering case. We see a similar $\sim 10\%$ change also at other detectors. Although the effect of collective neutrino oscillations, including fast-flavor conversions, is not yet clear, works indicate their promising occurrence (see, e.g., \cite{Nagakura:2023xhc}).

Potentially the largest error is the uncertainty on the failed supernova case. Because BH formation is very EOS dependent, the neutrino emission can vary dramatically between simulations. Since the PNS is continually accreting mass until BH formation time, the neutrino energies can become very high and result in a large enhancement of detectable DSNB neutrinos. This enhancement generally increases with $f_{\rm BH}$ (see Fig.~\ref{fig:bhfraction}). If the time to BH formation is very short, however, this can actually result in a reduction of events. Increased observations of failed supernovae, better constraints on the NS EOS, and precise measurement of the observed DSNB spectrum would be required to better understand the failed supernova channel (e.g., \cite{2018moller,Ashida:2022nnv}). 

In parallel to the theoretical modeling uncertainties, accurate estimation of the backgrounds will be necessary for detecting the DSNB at high significance. HK is expected to start taking data around $\sim2027$ which would add to the DSNB statistics of SK-Gd and JUNO. A similar analysis done here could be applied to HK as well since the signal could just be scaled up by the ratio of the volumes of the HK and SK-Gd detectors. However, we leave this for a future analysis after more is determined regarding efficiencies, background rates, and the addition of Gd. The background rates for DUNE are similarly not yet well known, but we do compute the estimated significance of DSNB detection at JUNO. We find that the time it takes to reach a significant detection is longer than what is computed for SK-Gd and JUNO in Ref.~\cite{Li:2022myd}. After folding in a catalog of new star formation rate density measurements and IMF-weighted neutrino emission spectra into our analysis, our signal turns out to be very similar to recent studies. However, because we have up-to-date data on the signal efficiency, the overall signal is decreased compared to previous studies. In the JUNO case, we do a rate-only analysis of significance, and an energy-bin-dependent analysis would increase the prospects, so this is conservative as well. Improved understanding of the backgrounds in these experiments is essential to detecting the DSNB significantly, alongside the reduction of theoretical uncertainties like SFRD measurements, the late phase, and failed supernovae. The fiducial signal rates of Tables~\ref{tab:eventserrors} and \ref{tab:ratesother} agree generally well with the computed rates with normal ordering compared to Refs.~\cite{2018moller,Sawatzki:2020mpb,Li:2022myd}.

In summary, we updated the inputs to the DSNB prediction in order to make more accurate estimates. Firstly, we improved on our understanding of the rate of core collapse by collating an up-to-date catalog of star formation rate measurements. This allows us to make estimates of the rate of core collapse without relying on a fit function, and allows us to quantify its uncertainty more directly from the measurements. We also use data from state-of-the-art CCSNe simulations to model the neutrino emission for the first $\sim5\,{\rm s}$ of 15 progenitors \cite{Nagakura:2021lma}. To estimate the late-phase neutrino emission after this over a timescale of $\sim10\,{\rm s}$, we use an analytic function fitted to the existing data and a method that correlates PNS mass and shock revival time to estimate the late-phase neutrino emission \cite{Ekanger:2022neg}. With these two models, we characterized the uncertainty in neutrino emission modeling. Finally, we used the data from several failed supernovae, including one 3D model, and existing $f_{\rm BH}$ estimates from observations to quantify the uncertainty that this unknown fraction has on DSNB rates.

Our fiducial predictions for the DSNB are $3.57\pm0.28^{+0.00+0.82}_{-1.58-1.94}$ events per year at SK-Gd with a flux of $5.10\pm0.4^{+0.00+0.50}_{-2.04-2.70}\,{\rm cm^{-2}~s^{-1}}$ between neutrino energy $9.3\,{\rm MeV}$ to $31.3\,{\rm MeV}$ (normal ordering), where the uncertainty comes from the late-phase neutrino emission treatment, SFRD, and $f_{\rm BH}$ errors, respectively. With a simplified rate analysis, we estimate that the DSNB is detectable at a level of $3\sigma$ by $\sim2030$ at SK-Gd and a level of $5\sigma$ by $\sim2035$ with a combined SK-Gd/JUNO analysis. After accounting for backgrounds, joint and long-term analyses will be crucial for detecting the DSNB and reducing the considerable uncertainty. Combined with steady improvements in CCSNe simulations, reduction of backgrounds, observations of CCSNe (successful and failed), and the advent of additional neutrino detectors, the prospect of detecting the DSNB in the next decade is extremely exciting.

\begin{acknowledgments}
We acknowledge David Vartanyan and Adam Burrows for providing Princeton's CCSNe models. We also thank Yusuke Koshio, Masayuki Harada, and Mark Vagins for comments and helpful discussions regarding the efficiency and backgrounds at SK-Gd. N. E. is supported by NSF Grants No.~AST1908960 and No. PHY-2209420. The work of S. H. is supported by the U.S.~Department of Energy Office of Science under Award No. DE-SC0020262, NSF Grants No.~AST1908960 and No.~PHY-2209420, and JSPS KAKENHI Grants No. JP22K03630 and No. JP23H04899, and the Julian Schwinger Foundation. H. N. is supported by Grant-inAid for Scientific Research (23K03468) and by the NINS International Research Exchange Support Program. This study was supported in part by World Premier International Research Center Initiative (WPI Initiative) by the Ministry of Education, Science and Culture of Japan (MEXT), by Grants-in-Aid for Scientific Research of the Japan Society for the Promotion of Science (JSPS, No.~JP22H01223), the MEXT (No.~JP17H06364, No. JP17H06365, No. JP19H05811, No. JP19K03837, No. JP20H01905), by the Central Research Institute of Explosive Stellar Phenomena (REISEP) at Fukuoka University and an associated Project No.~207002, and JICFuS as “Program for Promoting researches on the Supercomputer Fugaku” (Toward a unified view of the universe: from large scale structures to planets, JPMXP1020200109). S. R. was supported by a NSF REU Grant No.~PHY-1757087.

\end{acknowledgments}

\newpage

\appendix

\section{WEIGHTED-AVERAGE ESTIMATES}\label{sec:weightedavg}

Here we discuss how the rates, fluxes, and significance forecast changes if we assume the weighted-average over the simple average. The event rates and fluxes are presented in Table~\ref{tab:weightedavg}. Overall, the fiducial $R_{\nu}$ and $\phi$ are $\sim75\%$ of the simple average case. Since the signal reduces and not the background, the significance metric shown in Fig.~\ref{fig:significanceplot} decreases by roughly a factor of $\sqrt{75\%}\sim87\%$.

\begin{table}[h]
\caption{\label{tab:weightedavg}Table quantifying the fiducial signal and error of our DSNB signal rate, $R_{\nu}$, and integrated flux, $\phi$, at SK-Gd ($9.3 < E_{\nu} < 31.3\,{\rm MeV}$, where $E_{\nu}$ is the neutrino energy) for the inverse-error-weighted SFRD method. Values shown for both normal and inverted mass ordering, but calculated with the error-weighted averaging method. SFRD, LP, and BH errors calculated the same way as in Table~\ref{tab:eventserrors}.}
\begin{ruledtabular}
\begin{tabular}{cccccc}
&Ordering&Fiducial&SFRD err&LP err&BH err\\
\colrule
$R_{\nu}$&\textrm{NO}&2.63&$^{+0.07}_{-0.06}$&$^{+0.00}_{-1.16}$&$^{+0.61}_{-1.43}$\\
{[${\rm yr^{-1}}$]}&\textrm{IO}&2.08&$^{+0.07}_{-0.06}$&$^{+0.00}_{-0.79}$&$^{+0.13}_{-0.95}$\\
\hline
$\phi$&\textrm{NO}&3.75&$^{+0.10}_{-0.09}$&$^{+0.00}_{-1.56}$&$^{+0.39}_{-1.70}$\\
{[${\rm cm^{-2}~s^{-1}}$]}&\textrm{IO}&3.00&$^{+0.10}_{-0.09}$&$^{+0.00}_{-1.04}$&$^{+0.17}_{-1.39}$\\
\end{tabular}
\end{ruledtabular}
\end{table}

\section{UNOSCILLATED RESULTS}\label{sec:unoscillated}

In this section, we provide the yearly DSNB event rate and flux at SK-Gd for $\overline{\nu}_e$ and $\nu_x$, i.e. without assuming MSW oscillations, for the simple averaging method. One could then use these to estimate rates with a different oscillation consideration. The results are presented in Table~\ref{tab:unoscillated}.

\begin{table}[h]
\caption{\label{tab:unoscillated}Table quantifying the fiducial signal and error of our DSNB integrated flux, $\phi$, in the SK-Gd energy range ($9.3 < E_{\nu} < 31.3\,{\rm MeV}$, where $E_{\nu}$ is the neutrino energy). Values shown for $\nu_e$, $\overline{\nu}_e$, and $\nu_x$, unoscillated. SFRD, LP, and BH errors calculated the same way as in Table~\ref{tab:eventserrors}.}
\begin{ruledtabular}
\begin{tabular}{cccccc}
&Flavor&Fiducial&SFRD err&LP err&BH err\\
\colrule
$\phi$&$\nu_e$&4.50&$^{+0.40}_{-0.40}$&$^{+0.00}_{-1.95}$&$^{+1.49}_{-2.37}$\\
{[${\rm cm^{-2}~s^{-1}}$]}&$\overline{\nu}_e$&5.50&$^{+0.50}_{-0.50}$&$^{+0.00}_{-2.39}$&$^{+0.83}_{-2.79}$\\
&$\nu_x$&4.10&$^{+0.40}_{-0.40}$&$^{+0.00}_{-1.45}$&$^{+0.20}_{-1.92}$\\
\end{tabular}
\end{ruledtabular}
\end{table}

\bibliography{main}

\end{document}